%Paper: hep-th/9205016
%From: maharana@theory3.caltech.edu (J. Maharana)
%Date: Thu, 7 May 92 17:35:53 PDT

% Some macros for HEP preprints
\input phyzzx
%\hoffset=1truein
%\voffset=1.0truein
\hsize=6truein
\def\TITLEPAGE{\frontpagetrue}
\def\CALT#1{\hbox to\hsize{\tenpoint \baselineskip=12pt
	\hfil\vtop{\hbox{\strut CALT-68-#1}
	\hbox{\strut DOE RESEARCH AND}
	\hbox{\strut DEVELOPMENT REPORT}}}}

\def\CALTECH{\smallskip
	\address{California Institute of Technology, Pasadena, CA 91125}}
\def\TITLE#1{\vskip 1in \centerline{\fourteenpoint #1}}
\def\AUTHOR#1{\vskip .5in \centerline{#1}}

\def\ABSTRACT#1{\vskip .5in \vfil \centerline{\twelvepoint \bf Abstract}
	#1 \vfil}
\def\ENDTITLEPAGE{\vfil\eject\pageno=1}

\def\sqr#1#2{{\vcenter{\hrule height.#2pt
      \hbox{\vrule width.#2pt height#1pt \kern#1pt
        \vrule width.#2pt}
      \hrule height.#2pt}}}

\def\section#1#2{
\noindent\hbox{\hbox{\bf #1}\hskip 10pt\vtop{\hsize=5in
\baselineskip=12pt \noindent \bf #2 \hfil}\hfil}
\medskip}

\def\underwig#1{	% produce a tilde below the argument
	\setbox0=\hbox{\rm \strut}
	\hbox to 0pt{$#1$\hss} \lower \ht0 \hbox{\rm \char'176}}

\def\bunderwig#1{	% produce a tilde below the argument
	\setbox0=\hbox{\rm \strut}
	\hbox to 1.5pt{$#1$\hss} \lower 12.8pt
	 \hbox{\seventeenrm \char'176}\hbox to 2pt{\hfil}}

\def\MEMO#1#2#3#4#5{
\frontpagetrue
\centerline{\tencp INTEROFFICE MEMORANDUM}
\smallskip
\centerline{\bf CALIFORNIA INSTITUTE OF TECHNOLOGY}
\bigskip
\vtop{\tenpoint
\hbox to\hsize{\strut \hbox to .75in{\caps to:\hfil}\hbox to 3.8in{#1\hfil}
\quad\the\date\hfil}
\hbox to\hsize{\strut \hbox to.75in{\caps from:\hfil}\hbox to 3.5in{#2\hfil}
\hbox{{\caps ext-}#3\qquad{\caps m.c.\quad}#4}\hfil}
\hbox{\hbox to.75in{\caps subject:\hfil}\vtop{\parindent=0pt
\hsize=3.5in #5\hfil}}
\hbox{\strut\hfil}}}
\tolerance=10000
\hfuzz=5pt
\TITLEPAGE
\CALT{1781}
\TITLE {Duality and $O(d,d)$ Symmetries in String Theory\foot{Work supported in
part by the U.S. Dept. of Energy
under Contract no. DEAC-03-81ER40050.}}
\AUTHOR{Jnanadeva Maharana\foot{Permanent Address:  Institute of
Physics, Bhubaneswar - 752005, India.\hfil\break
\hskip -.2 truein email maharana@theory3.caltech.edu. until June 30, 1992.}}
\CALTECH
\ABSTRACT{The evolution of a closed bosonic string is envisaged
in the time-dependent background of its massless modes.  A duality
transformation is implemented on the spatial component of string
coordinates to obtain a dual string.  It is shown that the evolution
equations are manifestly $O(d,d)$ invariant.  The tree level string
effective actions for the original and the dual string theory are shown to
be equivalent.}

\ENDTITLEPAGE

\eject
It is now recognized that spacetime duality$^1$ plays an important role
in understanding several aspects of string theory.  One of the
consequences of $R$-duality is that the dynamics of a string on a circle
of radius $R$ is equivalent to that of another string on a circle of
radius ${1\over R}$  (in suitable units).  The idea of duality has also
been used in a wider context$^{2,3}$ and it is conjectured that this symmetry
might be maintained to all orders in string perturbation theory.$^4$
Furthermore, duality has been applied to study string cosmology$^{5-7}$
and used to obtain new black hole solutions.$^7$

Recently the concept of scale factor duality$^6$ (SFD) has been
introduced as a symmetry group of classical string equations of motion,
derived from a low energy string effective action.  One of the salient
features of SFD is that it does not require compactification of the target
space.  Moreover, this transformation relates different time-dependent
background configurations of string theory.  There is an intimate
connection between Narain's$^8$ construction of inequivalent static
compactification and the $O(d,d)$ transformations on background fields, which
rotate time-dependent backgrounds (solutions of equations of motion)
into other ones which are not necessarily equivalent.  Subsequently,
several cosmological$^9$ and black hole solutions$^{10}$ have been
obtained through the implementation of $O(d,d)$ transformations and
their generalizations.  It has been observed that the roles of the
canonical momentum $P$ and $X'$ are interchanged under duality (this
amounts to interchange of winding number and momentum zero modes) when
constant background fields are suitably transformed along with
$P\leftrightarrow X'$.  The Hamiltonian remains invariant under duality.

The purpose of this note is to investigate the evolutions of a closed
bosonic string in a background in which background of its massless
excitations (graviton, $g_{\mu\nu}$, antisymmetric tensor, $B_{\mu\nu}$,
and dilaton $\phi$) are time-dependent. It is shown that the evolution
equations of the string in a time-dependent background have a hidden
$O(d,d)$ symmetry; $d$ is the number of space
dimensions.  The space-time dimensions is $D = d + 1$.  We introduce a
duality transformation on string coordinates to define a dual
Lagrangian.  Then a larger manifold is constructed to include string
coordinates and their dual coordinates.  The equations of motion are
derived in a manifestly $O(d,d)$ invariant manner.  It is worthwhile to
mention that although the equations of motion are manifestly $O(d,d)$
invariant the Lagrangian is not.  We may recall that a similar situation
also arises in the context of the discussion of noncompact hidden
symmetries in supergravity theories,$^{11}$ where the equations of
motion are
manifestly invariant under these hidden symmetries whereas the action is
not.

It is argued that the vanishing $\beta$-function equations of the
original string theory are the same as those of the dual string theory,
since the tree level string effective action for both the theories are
the same when the background are time-dependent (but not space-dependent)

The two-dimensional sigma model Lagrangian that we consider is
$$\eqalignno{\bar L &= {1\over 2} \sqrt{-\gamma} \gamma^{ab} \partial_a X^\mu
\partial_b X^\nu g_{\mu\nu} + {1\over 2} \epsilon^{ab} \partial_a X^\mu
\partial_b X^\nu B_{\mu\nu}\,\, ,\cr
&+ L_D\,\, ,&(1)\cr}$$
where $\gamma_{ab}$ is the world sheet metric; $\epsilon^{ab}$ the
antisymmetric tensor, such that $\epsilon^{01} =  1, \mu,\nu$ are
$D$-dimensional target space indices and $a,b$ are world sheet indices
respectively.  $L_D$ describes the coupling of a dilaton background with
string, whose explicit form will be discussed later.  All the background
fields $g_{\mu\nu}, B_{\mu\nu}$ and $\phi$ are allowed to depend on the
time coordinate $X^0$ only.  It is convenient to bring $g_{\mu\nu}$ and
$B_{\mu\nu}$ to the following special form by implementing general
coordinate transformation and Abelian gauge transformation respectively.
$$g_{\mu\nu} = \left[\matrix{-1 & 0\cr 0 & G_{ij}(t)\cr}\right] ~~{\rm
and}~~ B_{\mu\nu} = \left[\matrix{0 & 0\cr 0 & B_{ij}(t)\cr}\right]\,\,
,
\eqno (2)$$
$i,j = 1, d$ are indices of the spatial coordinates.  Then the
Lagrangian $\bar L$ can be re-expressed as
$$\eqalignno{\bar L &= - {1\over 2} \sqrt{-\gamma} \gamma^{ab} \partial_a X^0
\partial_b X^0 + {1\over 2} \sqrt{-\gamma} \gamma^{ab} \partial_a X^i
\partial_b X^j G_{ij}\cr
& + {1\over 2} \epsilon^{ab} \partial_a X^i \partial_b X^j B_{ij} +
L_D\,\, . &(3)\cr}$$
The equations of motion,
$$	{\partial\bar L\over \partial X^\mu} - \partial_a {\partial\bar
L\over \partial \partial_a X^\mu} = 0\,\, , \eqno (4)$$
take the following form for the spatial components of the string
coordinates $\{X^i\}$:
$$\partial_a {\cal A}_i^a = 0\,\, , \eqno (5)$$
with
$${\cal A}_i^a \equiv {\partial\bar L\over\partial\partial_a X^i}
= \sqrt{-\gamma} \gamma^{ab} \partial_b X^j G_{ij} + \epsilon^{ab}
\partial_b X^j B_{ij}\,\, , \eqno (6)$$
since there is no explicit $X^i$ dependence in $\bar L$.  However, the
equation of motion for the $X^0$ coordinate is much more complicated.
$${1\over 2} \sqrt{-\gamma} \gamma^{ab} \partial_a X^i
\partial_b X^j {\partial\over\partial X^0} G_{ij} + {1\over 2}
\epsilon^{ab} \partial_a X^i \partial_b X^j {\partial\over\partial X^0}
B_{ij}$$
$$+ \partial_a (\sqrt{-\gamma} \gamma^{ab} \partial_b X^0) +
{\partial L_D\over \partial X^0} - \partial_a \left({\partial L_D\over
\partial \partial_a X^0}\right) = 0\,\, . \eqno (7)$$
Note that (7) contains terms with derivatives with respect to $X^0$.
Therefore, it cannot be written in the form of eq. (5).  In the BRST
quantization of string theory it is necessary to introduce ghost fields
in the gauge fixed action.  In this framework in stead of the usual
dilaton coupling
$$	\int d^2 \sigma \sqrt{-\gamma} R^{(2)} \phi (X)\,\,, \eqno (8)$$
the dilation is coupled to the ghost current$^{12,13}$ (when ON gauge
fixing is adopted).
$$	L_D = {1\over 2} \int d^2 \sigma [\bar\partial \phi (X) b_{++}
c^+ + \partial \phi b_{--} c^-]\,\, ,\eqno (9)$$
where $\phi$ is the dilaton and $b_{\pm\pm}$ and $c^\pm$ are the ghost
fields.  Notice that since we assume that $\phi$ depends on $X^0$ only,
$\bar\partial \phi (X) = {\partial\over\partial X^0} \phi(X^0)
\bar\partial X^0$ and $\partial\phi (X) = {\partial\over\partial X^0}
\phi (X^0) \partial X^0$.  In what follows we shall utilize the property
of $\bar L$ given by (1) that the equations of motion for $\{X^i\}$ is a
divergence (5).  Thus we are led to construct a new Lagrangian, $L_1$,
in the first order formalism where we deal with $\{X^i\}$
coordinates.  Introducing a field $u_a^i$, we write
$$	L_1 = - {1\over 2} \sqrt{-\gamma} \gamma^{ab} u_a^i u_b^j G_{ij}
- {1\over 2} \epsilon^{ab} u_a^i u_b^j B_{ij}$$
$$	+ \partial_a X^i (\sqrt{-\gamma} \gamma^{ab} u_b^j G_{ij} +
\epsilon^{ab} u_b^j B_{ij})\,\, . \eqno (10)$$
The $u_a^i$ variation of $L_1$ gives
$$	{\partial L_1\over \partial u_a^i} = (\partial_b X^j -
u_b^j)(\sqrt{-\gamma} \gamma^{ab} G_{ij} + \epsilon^{ab} B_{ij}) = 0
\,\, , \eqno (11)$$
as the $u_a^i$ equations of motion since $L_1$ does not contain any
derivative of the field; whereas $\partial_a X^i$ variation gives
$$	{\partial L_1\over\partial \partial_a X^i} = \sqrt{-\gamma}
\gamma^{ab} u_b^j G_{ij} + \epsilon^{ab} u_b^j B_{ij} \,\, , \eqno
(12)$$
with the equations of motion $\partial_a \left({\partial
L_1\over\partial \partial_a X^i}\right) = 0$.  If we solve for
$\partial_a X^i = u_a^i$ from (10) and substitute in (8), we receive
the expression for ${\cal A}_i^a$, eq. (5) and the equations of motion
for $\{X^i\}$.

Let us consider another first order Lagrangian $L_2$ with $a$ variables
$Y_i$, and an auxiliary field (again denoted by $u_a^{i}$)
$$	L_2 = {1\over 2} \sqrt{-\gamma} \gamma^{ab} u_a^i u_b^j G_{ij} +
{1\over 2} \epsilon^{ab} u_a^i u_b^j B_{ij}$$
$$	+ \epsilon^{ab} \partial_a Y_i u_b^j \,\, . \eqno (13)$$
Variation of $L_2$ with respect to $u_a^{i}$ gives the relation
$$	\epsilon^{ab} \partial_b Y_i = \sqrt{-\gamma} \gamma^{ab} u_b^j
G_{ij} + \epsilon^{ab} u_b^j B_{ij} \,\, , \eqno (14)$$
and the $Y_i$ equation of motion is
$$	\partial_a \left({\partial L_2\over \partial \partial_a
Y_i}\right) = \partial_a \left(\epsilon^{ab} u_b^{i}\right) = 0 \,\, . \eqno
(15)$$
Solving for $u_i^a$ in terms of $Y_{i}$ in (14) gives us
$$u_a^i = {1\over\sqrt{-\gamma}} \gamma_{ab} \epsilon^{bc} {\bf A}^{ij}
\partial_c Y_j + {\bf F}^{ij} \partial_a Y_j \,\, , \eqno (16)$$
where ${\bf A} = B^{-1} (GB^{-1} - BG^{-1})^{-1}$ and ${\bf F} = -
G^{-1} (GB^{-1} - BG^{-1})^{-1}$ are symmetric and antisymmetric
time-dependent matrices respectively as is evident from the symmetry
properties of backgrounds $G$ and $B$.

The equations of motion derived from $L_1$ suggests that we can unite
$u_i^a = \epsilon^{ab} \partial_b Y_i$ locally; whereas the $Y_i$
equation of motion derived from $L_2$ allows us to write ${\partial
L_2\over \partial \partial_a Y_i} = \epsilon^{ab} \partial_b X^i$.  Thus
the $\partial_a X^i$ and $\partial_a Y_i$ variations of $L_1$ and $L_2$
(after substituting the auxiliary fields) locally take the form
$$	\epsilon^{ab} \partial_b Y_i = {\partial L_1\over\partial
\partial_a X^i} = \sqrt{-\gamma} \gamma^{ab} \partial_b X^j G_{ij} +
\epsilon^{ab} \partial_b X^j B_{ij} \eqno (17)$$
$$	\epsilon^{ab} \partial_b X^i = {\partial L_2\over \partial
\partial_a Y_i} = \sqrt{-\gamma} \gamma^{ab} \partial_b Y_j {\bf A}^{ij}
+ \epsilon^{ab} \partial_b Y_j {\bf F}^{ij}\,\, . \eqno (18)$$
As always happens with such dual reformulation, the field equations
derived for the string coordinates $\{X^i\}$ are the Bianchi identities
for the dual variables $\{Y_i\}$, whereas the equation of motion for
$\{Y_i\}$ are the Bianchi identities for $\{X^i\}$.  Moreover, the
matrices ${\bf A}$ and ${\bf F}$ play the role of metric and
antisymmetric tensor field for the dual string coordinates.  It follows
from (17) and (18) that the canonical momenta $\{P_i\}$ of $\{X^i\}$ are
identified with $Y_i'$ whereas those of $\{Y_i\}$ are with $\{X^{'i}\}$.  We
may recall that, for constant background fields, the role of $P$ and
$X'$ are interchanged under duality.  In order to reveal the hidden symmetries
associated with the equations of  motion, we enlarge the manifold where
$X^i$ and $Y_i$ are treated as independent coordinates (this is
analogous to the phase space in a Hamiltonian formulation of dynamics).
Let us first rewrite equations (17) and (18) as
$$	\sqrt{-\gamma} \epsilon_{ab} \gamma^{bc} \partial_c X^i = G^{ij}
\partial_a Y_i - G^{ij} B_{jk} \partial_a X^k \eqno (19)$$
$$	\sqrt{-\gamma} \epsilon_{ab} \gamma^{bc} \partial_c Y_i = ({\bf
A}^{-1})_{ij} \partial_a X^j - ({\bf A}^{-1})_{ij} F^{jk}
\partial_a Y_k \,\, . \eqno (20)$$
Let $W$ denote the $2d$ coordinates $\{X^i, Y_i\}$ collectively; then
eqs. (19) and (20) can be written in a compact form as the single
equation
$$	{\bf M} \eta \partial_a W = \sqrt{-\gamma} \epsilon_{ab}
\gamma^{bc} \partial_c W \,\, , \eqno (21)$$
where the symmetric $2d \times 2d$ matrix
$$	{\bf M} = \left(\matrix{G^{-1} & - G^{-1} B\cr BG^{-1} & G -
BG^{-1} B\cr}\right) \,\, \eqno (22)$$
is the same one as appears in the discussion of the duality properties
of string effective action,$^9$ and
$$	\eta = \left(\matrix{0 & {\bf 1}\cr {\bf 1} & 0\cr}\right) \,\,
, \eqno (23)$$
is the $O(d,d)$ metric, ${\bf 1 }$ being $d \times d$ unit matrix.  It
is easy to check that
$$	{\bf M} \eta {\bf M} = \eta ~~{\rm and}~~ \eta {\bf M} \eta =
{\bf M}^{-1} \,\, . \eqno (24)$$
Thus we conclude that ${\bf M} ~\epsilon~ O(d,d)$.  Equation (21) can be
rewritten as
$$	\epsilon^{ab} \partial_b W = \eta {\bf M}^{-1} \sqrt{-\gamma}
\gamma^{ab} \partial_b W$$
leading to the manifest $O(d,d)$ invariant integrability equation
$$	\partial_a (\eta {\bf M}^{-1} \sqrt{-\gamma} \gamma^{ab}
\partial_b W) = 0 \,\, . \eqno (25)$$
We note that the equations of motion are invariant under $O(d,d)$
transformations although the action is not invariant under duality
transformation.  This is one of the characteristics of hidden symmetries
associated with duality transformations as has been emphasized by
Gaillard and Zumino.$^{11}$

Let us turn our attention to the conformal invariance of these theories
and the equations of motion satisfied by the background fields.  The
Hamiltonian associated with $L_1$ together with the contributions of the
$X^0$ coordinates and the dilaton term is
$$	H = {1\over 2} (P_0 + X^{'0})^2 + {1\over 2} (P~ X') {\bf M} (P~ X')^T +
H_D \,\, , \eqno (26)$$
where the first term is the contribution of the $X^0$ coordinates, $P_0$
being momentum conjugate to $X^0$ whereas $\{P_i\}$ are the conjugate
momentum of $\{X^i\}$ and ${\bf M}$ is the matrix defined in eq. (22).
$H_D$ is the Hamiltonian associated with
the dilaton coupling to the string.$^{13}$  The other constraint, which
generates $\sigma$-reparametrization, can be written as
$$	P_0 X^{'0} + {1\over 2} (P~ X') \eta (P~ X')^T \,\, . \eqno (27)$$
If we now demand conformal invariance of the theory we derive the
equations of motion for the background fields which  ensure the
vanishing of the associated $\beta$-functions.  These conditions can
also be derived from the variation of the tree level string effective
action.  Indeed, such an effective action has been obtained in a
compact form for time-dependent background fields by Meissner and
Veneziano$^9$
$$	S_E = \int dt e^{-\varphi} [\Lambda + \dot \varphi^2 + {1\over
8} Tr (\partial_t {\bf M} \eta \partial_t {\bf M} \eta)] \,\, , \eqno
(28)$$
where $\varphi = \phi - \ell n \sqrt{\det G}$, is the shifted dilaton, $G$ is
as defined in eq. (2), and $\Lambda$ is the cosmological term
proportional to (D - 26) that appears for a noncritical bosonic string.
${\bf M}$ and $\eta$ are defined in eqs. (22) and (23).

Let us now implement the duality transformation.  Note that the $X^0$
coordinate and $H_D$ remain unaffected since $\{X^i\}$ transform to
$\{Y_i\}$ under duality.  Thus the generator of
$\sigma$-reparametrization transformation has the form
$$P_0 X^{'0} + {1\over 2} (\tilde{P}~ Y') \eta (\tilde{P}~ Y')^T \,\, , \eqno
(29)$$
and the dual Hamiltonian is
$$\tilde{H} = {1\over 2} (P_0 + X^{'0})^2 + {1\over 2} (\tilde{P}~ Y')
\tilde{{\bf M}} (\tilde{P}~ Y')^T + H_D \,\, , \eqno (30)$$
where $\{\tilde{P}^i\}$ are conjugate momenta of $\{Y_i\}$.  $\tilde{{\bf
M}}$ can be computed to be
$$	\tilde{{\bf M}} = \left[\matrix{G - BG^{-1} B & BG^{-1}\cr
-G^{-1} B & G^{-1}\cr}\right] = {\bf M}^{-1} \,\, . \eqno (31)$$
It is easy to write down the Meissner-Veneziano$^9$ effective action for the
dual theory with $\tilde{{\bf M}}$ and it reads ($\varphi$ remains
unchanged)
$$	\tilde{S}_E = \int dt e^{-\varphi} \left[\Lambda + \dot \varphi^2 +
{1\over 8} Tr (\partial_t \tilde{{\bf M}} \eta \partial_t \tilde{{\bf
M}} \eta) \right.\,\, . \eqno (32)$$
It follows from the properties of ${\bf M}$ eq. (24) and $\eta^2 = 1$
that
$$\tilde{S}_E = S_E \,\, . \eqno (33)$$
Therefore, the tree level string effective actions for the two theories
are the same.

It is natural to ask what symmetries the string evolution equations will
exhibit if the background fields depend on some of the spatial
coordinates in addition to time.  For example, one could envisage a
situation where background fields are independent of coordinates
$X^\alpha, \alpha = 1, .. m, m < D - 1$.  It has been shown by
Sen,$^{14}$ in the framework of string field theory, that the string
effective action in this case has an $O(m) \otimes O (m)$ symmetry.
However, in our approach, it is not easy to demonstrate such invariance
properties of the string equations of motion, since it is not possible
to  transform the metric and antisymmetric tensor to a simple form as in
eq. (2) in the general case.  It might be possible to show the
invariance of the string equations of motion by introducing a more
general duality transformation than the one used here.

To summarize: we considered a closed bosonic string in the time-dependent
background of its massless modes.  A duality transformation
involving only spatial string coordinates was utilized to obtain a new
Lagrangian.  Then it was shown that the string evolution equations are
$O(d,d)$ invariant.  It was argued that the tree  level string effective
action and the dual effective  action are equivalent.  Therefore, the
$\beta$-functions associated with the original string theory and the
dual theory are identical at least to lowest order in $\alpha'$.

I would like to thank John Schwarz for valuable discussions and for
critically reading the manuscript, and to Keke Li for useful comments.
It is a pleasure to acknowledge the gracious hospitality of the High Energy
Physics group, especially of John Schwarz, at Caltech.

\noindent {\bf References}

\item{1.}  K. Kikkawa and M. Yamasaki, Phys. Lett. {\bf B149} (1984)
357; N. Sakai and I. Senda, Prog. Th. Phys. {\bf 75} (1986) 692.  For a
recent review see ``Spacetime duality in string theory,'' J.H. Schwarz
in Elementary Particles and the Universe, Essays in Honor of Murray
Gell-Mann, ed. J.H. Schwarz, Cambridge University Press 1991 and
references therein, and Caltech Preprint CALT-68-1740, May 1991.

\item{2.}  V.P. Nair, A. Shapere, A. Strominger and F. Wilczek, Nucl.
Phys. {\bf B287} (1987) 402; A. Shapere and F. Wilczek, Nucl. Phys.
{\bf B320} (1989) 669; T.H. Buscher, Phys. Lett. {\bf B194} (1987) 59
and {\bf 201} (1988) 466; A. Giveon, E. Ravinovici and G. Veneziano,
Nucl. Phys. {\bf B322} (1989) 169.

\item{3.}  S. Celcotti, S. Ferrara and L. Girardello, Nucl. Phys. {\bf
B308} (1988) 436; P. Ginsparg and C. Vafa, Nucl. Phys. {\bf B289} (1987)
414; M. Dine, P. Huet and N. Seiberg, Nucl. Phys. {\bf B322} (1989) 301;
J. Lauer, J. Mas and H.P. Nilles, Phys. Lett. {\bf B226} (1989) 251; W.
Lerche, D. L\"ust and N. Warner, Phys. Lett. {\bf B231} (1989) 417; B.R.
Greene and M.R. Plesser, Nucl. Phys. {\bf B338} (1990) 15; M.J. Duff,
Nucl. Phys. {\bf B335} (1990) 610; J.M. Molew and B. Ovrut, Phys. Rev.
{\bf D40} (1989) 1146; T. Banks, M. Dine, H. Dijrstra and W. Fischler,
Phys. Lett. {\bf B212} (1988) 45.

\item{4.}  E. Alvarez and M. Osorio, Phys. Rev. {\bf D401} (1989) 1150.

\item{5.}  R. Brandenberger and C. Vafa, Nucl. Phys. {\bf B316} (1988)
391; A.A. Tseytlin, Mod. Phys. Lett. {\bf 6A} (1991) 1721; A.A. Tseytlin and
C. Vafa, Nucl. Phys. {\bf B372} (1992) 443.

\item{6.}  G. Veneziano, Phys. Lett. {\bf B205} (1991) 287.

\item{7.}  A. Giveon,
LBL Preprint 30671, E. Smith and J. Polchinski, Preprint UTTG-07-91; M.
Rocek and E. Verlinde, IAS Preprint IASNS-HEP-91/68, ITP-SB-91-53; A.
Giveon and M. Rocek, IAS Preprint IASNS-HEP-91/84, ITP-SB-91-67.

\item{8.}  K.S. Narain, Phys. Lett. {\bf B169} (1986) 61; K.S. Narain,
H. Sarmadi and E. Witten, Nucl. Phys. {\bf B279} (1987) 309.

\item{9.}  K.A. Meissner and G. Veneziano, Phys. Lett. {\bf B267} (1991)
33; K.A. Meissner and G. Veneziano, CERN Preprint CERN-TH-6235/91;
M. Gasperini, J. Maharana and G. Veneziano, Phys. Lett. {\bf B272} (1991)
167; M. Gasperini and G. Veneziano, Phys. Lett. {\bf B277} (1992) 256;
J. Panvel, Preprint LTH 282 DAMTP, University of Liverpool.

\item{10.}  A. Sen, Phys. Lett. {\bf B272} (1992) 34; F. Hassan and A. Sen,
TIFR Preprint TIFR/TH/91-40; S.P. Khastgir and A. Kumar, Mod. Phys. Lett.
{\bf 6A} (1991) 3365.

\item{11.}  S. Ferrara, J. Scherk and B. Zumino, Nucl. Phys. {\bf B121}
(1977) 393; E. Cremmer, J. Scherk and S. Ferrara, Phys. Lett. {\bf B68}
(1977) 234; E. Cremmer and B. Julia, Nucl. Phys. {\bf B159} (1979) 149;
M.K. Gaillard and B. Zumino, Nucl. Phys. {\bf B193} (1981) 221.

\item{12.}  T. Banks, D. Nemeschansky and A. Sen, Nucl. Phys. {\bf B277}
(1986) 67.

\item{13.}  J. Maharana and G. Veneziano, Nucl. Phys. {\bf B283} (1987)
126.

\item{14.}  A. Sen, Phys. Lett. {\bf B271} (1991) 295.

\bye